\documentstyle[12pt, psfig]{elsart}

\begin{document}

\voffset = 1.0truecm

\newcommand{\solar}{\ifmmode_{\mathord\odot}\else$_{\mathord\odot}$\fi}
\newcommand{\epr}{e-print  astro-ph/  } 
\newcommand{\gray}{ $\gamma$-ray}
\newcommand{\grays}{ $\gamma$-rays}
\newcommand{\apj}{ Astrophys. J.}
\newcommand{\prl}{ Phys. Rev. Letters}%\newcommand{\etal}{ {\it et al.} }

\begin{frontmatter}
\title{Are Diffuse High Energy Neutrinos and \grays\ from Starburst Galaxies 
Observable?}

\author{F.W. Stecker}

\address{NASA Goddard Space Flight Center, Greenbelt, MD, USA}

\begin{abstract}

Loeb and Waxman have argued that high energy neutrinos from the
decay of pions produced in interactions of cosmic rays with  
interstellar gas in starburst galaxies would be produced with
a large enough flux to be observable. Their model is
reexamined here and we obtain an upper limit to the diffuse neutrino
flux from starburst galaxies. 
The upper limit obtained here is a factor of $\sim$5 lower than
the flux which they predict. Our predicted neutrino flux would be 
below the atmospheric neutrino foreground flux at energies below
$\sim$ 300 TeV and therefore would be unobservable.
Compared with predicted fluxes from other extragalactic high energy
neutrino sources, starburst neutrinos with $\sim$ PeV energies
would have a flux considerably below that predicted for AGN models.

We also estimate an upper limit for the diffuse GeV \gray\ flux from
starbust galaxies to be $\cal{O}$$(10^{-2})$ of the observed \gray\
background, much less than the component from unresolved blazars and
more than an order of magnitude below the estimate of Thompson {\it \etal}

\end{abstract}


\end{frontmatter}

\section{Introduction}

Interactions of cosmic-ray nuclei with interstellar gas nuclei 
in our galaxy produce secondary pions. The neutral pions then 
decay to produce most of the galactic $\gamma$-rays above 0.1 GeV 
which have been observed~\cite{st77}; the decay of the charged 
pions produces galactic cosmic-ray neutrinos~\cite{st79}.

It was pointed out 30 years ago that the distribution of high
energy \grays\ in our Galaxy is related to the distribution
of molecular clouds and very young hot high-mass stars in OB 
associations which are short-lived and explode into 
supernovae~\cite{st75,st76}.
This association between supernovae which are likely to produce 
cosmic rays and dense regions of molecular gas led to the scenario
where interactions between the gas and cosmic rays then
produce the \grays\ {\it via} the decay of the $\pi^0$ mesons produced
in these interactions. A natural implication then would be that
galaxies which were undergoing a phase of extremely active star 
formation would be likely sources of high energy \grays. 
Such sources are known as ``starburst galaxies''. 
Detailed model calculations have been made to predict \gray\
fluxes above 100 MeV for the starburst galaxies Arp 220~\cite{to04}
and NGC 253~\cite{to05}. The predicted flux for Arp 220 is close
to the sensitivity limit for the {\it GLAST} large area 
telescope whereas the predicted flux for NGC 253 is an order of
magnitude higher than that sensitivity limit\footnote{For information 
on {\it GLAST} see {\tt http://www-glast.stanford.edu}} and close to
the upper limit obtained from the {\it EGRET} data~\cite{bl99}.

Loeb and Waxman (LW) have suggested that such hadronic processes 
in starbust galaxies,
involving cosmic ray interactions with interstellar gas followed by
the decay of $\pi^{\pm}$'s, can produce, {\it in toto},
a large enough background of diffuse high energy neutrinos to
be observable~\cite{lw06} with a very large neutrino detector
such as {\it Icecube}~\cite{ha06}. LW then argue 
that radio observations of starburst galaxies imply a {\it lower limit} on
the cumulative extragalactic neutrino flux from starburst galaxies 
which is within the sensitivity range of {\it Icecube}. This argument is
examined here and the opposite conclusion is obtained. Indeed, we derive an
{\it upper limit} to the cumulative high energy neutrino flux from
starburst galaxies. The diffuse flux of GeV \grays\ from the same processes
is found to be $\cal{O}$$(10^{-2})$ of the observed \gray\
background, much less than the component from unresolved blazars and
more than an order of magnitude below the estimate of Thompson {\it \etal}
~\cite{th06}.

\section{Radio Emission and the Neutrino Flux from Starbust Galaxies}

LW start with the observed synchrotron
emission from starburst galaxies which is produced by relativistic
electrons in these sources~\cite{yu01}. They then make the assumptions
that (1) the presence of relativistic electrons in these sources implies 
the presence of relativistic protons, (2) the protons lose essentially
all of their energy to pion production, and (3), a lower limit to the
energy loss rate of the protons can then be obtained from the synchrotron
radio flux by assuming that all of the electrons (and positrons) which
are radiating are from pion decay. 

Assumption (1) is a reasonable one which is supported by observations
of cosmic rays in our own Milky Way galaxy. Assumption (3), {\it viz.}, the
``lower limit'' assumption depends on assumption (2). However,
assumptions (2) and (3) can be questioned because (a) the synchrotron radiating
electrons may be largely accelerated primaries rather than secondaries
related to pion production and decay as is the case in our own Galaxy,
and (b) the conditions in starburst
galaxies are significantly different from those in our own galaxy.
In particular, starburst galaxies exhibit strong ``superwinds''~\cite{ho03}. 
Such winds have significant dynamical effects and may disrupt
magnetic fields and drive
protons out of these galaxies before they can lose all of their energy
by interacting with interstellar gas nuclei to produce pions.
In contrast, assumption (2) of LW assumes full trapping of relativistic
nuclei in the disks of starburst galaxies to the point where they only
lose energy in hadronic interactions with gas atoms. This is in stark 
contrast to the situation in our own galaxy. (See footnote 2.)

These two caveats call into serious question the argument that the
radio data can provide a true lower limit on the cumulative
diffuse flux of neutrinos from starburst galaxies. Let us, however,
ignore them and consider that assumptions (1)-(3) are reasonable for 
obtaining an analytic {\it upper limit} for such a flux. Let us then
accept the other estimates which lead to the ratio of injected power
of protons to electrons at a fixed particle energy, $\eta_{p/e} \simeq 6$
and a neutrino luminosity which is then related to the local radio 
luminosity density by 

\begin{equation}
E_{\nu}^2\Phi_{\nu} (E_{\nu} = 1 GeV) \simeq (ct_{H}/4\pi)\zeta
[4f(dL_{f}/dV)]_{f = 1.4 GHz}
\end{equation}

\noindent where $t_{H}$ is the age of the universe
and $\zeta = 3$ is an evolution factor which takes account of the fact
that starburst galaxies were more numerous in the past~\cite{lw06}.
 LW take the local energy production rate per unit 
volume at a frequency f = 1.4 GHz to be $\simeq 10^{28.5}$ W Mpc$^{-3}$.
Let us reexamine this value for $f(dL_{f}/dV)]_{f = 1.4 GHz}$.

The local 1.4 GHz energy production rate has been derived by LW
by making use of an important connection bewteen radio emission and far
infrared (FIR) emission in galaxies given in the paper of 
Yun, Reddy and Condon (YRC)~\cite{yu01}. That paper uses the data on
{\it IRAS} galaxies to derive the local infrared luminosity density at 
60 $\mu$m to be $2.6 \times 10^7 L_{\solar}$ Mpc$^{-3}$. This {\it total}
power density is then used by LW to obtain the 1.4 GHz power density {\it via} 
a strong empirical correlation between the FIR and 1.4 GHz 
luminosity densities.

The key difference between the result to be derived here and that obtained
by LW is in chosing how to interpret the paper of YRC.
YRC state that less than 10\% of the local FIR luminosity density 
is contributed by luminous IR galaxies with $L_{FIR} > 10^{11} M_{\solar}$;
this is the component which includes the starburst galaxies. (Figure
11 of YRC yields an estimate of $\sim$ 6\%.)

If we take the local contribution from starburst galaxies at 60$\mu$m to 
be $2.6 \times 10^6L_{\solar}$ (the 10\% upper  limit found by YRC),
using the relation (4) of YRC as shown in their Fig. 5(a) one
finds that the component of the local radio luminosity density related
to   the    starburst   galaxies   is   at    most   $\Phi_{\rm 1.4
GHz} = 10^{18.4}$ W Hz$^{-1}$Mpc$^{-3}$ which, when  multiplied by 1.4  
GHz, gives $\Phi_{1.4  GHz} <  10^{27.5}$ W  Mpc$^{-3}$. This value 
is an order  of  magnitude  lower  than   the  flux  obtained  
by  LW~\cite{lw06}.\footnote{One might also ask why not consider
more ``normal'' galaxies with lower FIR luminosities and add them all in to
estimate a higher neutrino flux? We note that in the case of our own
galaxy, cosmic rays lose only a small fraction of their energy in
pion producing interactions before escaping the disk, contrary to
assumption (2). Also, cosmic rays with energies above 1 PeV
(the relevant range for producing 100 TeV neutrinos) have a differential
particle spectrum $\propto E^{-3.2}$, much steeper than the $E^{-2}$
spectrum assumed by LW. This spectrum probably reflects
both a steeper composite source spectrum and a shorter confinement time than
those at lower energies.}

This is not the whole story because there is a higher relative fraction of the
energy input from the higher relative number of starburst galaxies at higher 
redshifts. To estimate this effect, we assume that the FIR background is 
proportional to the integrated star formation activity rate. The fraction of 
the FIR background, $\kappa(\Delta z)$ contributed by galaxies in different 
redshift ranges, $\Delta z$, is obtained from Ref.~\cite{la05}. Then we 
multiply $\kappa(\Delta z$) by the fraction of the FIR background contributed 
by starburst galaxies in different redshift ranges, $\xi(\Delta z)$, 
to estimate the mean fraction of the total FIR background contributed by 
starburst galaxies. Estimates for $\kappa$ and $\xi$ are shown in Table 1.

\begin{table*}[ht]
\vspace{0.6cm}
\centerline{Table 1: Relative contributions to the $\nu$ Starburst Galaxy Flux (see text).} 
\vspace{0.5cm} 
\begin{center}
\begin{tabular}{|cccc|}  \hline \hline

Redshift Range ($\Delta z$) & $\kappa(\Delta z)$~\cite{la05} & $\xi(\Delta z)$ & Reference for $\xi$ \\
\hline
0 to 0.2 & 10\% & $<$ 10\% & \cite{yu01} \\
0.2 to 1.2 & 68\% & $\sim$ 13\% & \cite{la05}
 \\
$>$1.2  & 22\% & $\sim$ 60\% &  \cite{er06} \\

\hline

\end{tabular}
\end{center}
\vspace{0.6cm}
\end{table*}

Using the results from Table 1, we estimate that 23\% of the observed
FIR background integrated over redshift is from starburst galaxies.

\section{Observability of High Energy Neutrinos from Starburst Galaxies}

The upper limit on the radio flux from starburst galaxies obtained above can 
be used to obtain an upper limit on the neutrino flux from starburst
galaxies by using equation (1) as derived by LW. One then finds that the  
neutrino background
energy  flux from starburst  galaxies 
would be at most $\sim 2 \times  10^{-8}$ GeV cm$^{-2}$ s$^{-1}$ sr$^{-1}$ 
Such a flux  would be undetectable above the atmospheric background
neutrino flux, even  if  equation (1)  is  assumed to  be 
valid  when extrapolated to 300 TeV and  even granting all of the 
assumptions made by LW.

\begin{table*}[ht]
\centerline{Table 2: Neutrino Energy Fluxes (GeV cm$^{-2}$ s$^{-1}$ sr$^{-1}$)}
\begin{center}
\begin{tabular}{|ccccc|}  \hline \hline

$\nu$ Source &  $E^2\Phi(10 \rm TeV)$ &  $E^2\Phi(100\rm TeV)$ & $E^2\Phi(1 \rm PeV)$ & Reference \\
\hline 
Atm: AMANDA-II & $2 \times 10^{-6}$ & $7 \times 10^{-8}$ & $<3 \times 10^{-9}$ & \cite{bo06} \\
Atm (Vertical) & $7 \times 10^{-7}$ & $\sim 2 \times 10^{-9}$  & --- & \cite{ga05} \\ 
AMANDA-II Diff.Lim. & $9 \times 10^{-8}$ & $9 \times 10^{-8}$ & $9 \times 10^{-8}$ & \cite{hi06} \\ \hline

Starburst Galaxies & $< 2 \times 10^{-8}$ & $< 2 \times 10^{-8}$ & $< 2 \times 10^{-8}$ & This paper \\ 
AGN Cores &   $5 \times 10^{-10}$ & $10^{-8}$ & $10^{-7}$  &\cite{st05} \\ 
AGN & $3 \times 10^{-9}$ & $3 \times 10^{-8}$ & $2 \times 10^{-7}$ & \cite{mpr} \\ 
GRB & $5 \times 10^{-10}$ & $3 \times 10^{-9}$ & $3 \times 10^{-9}$ & \cite{wb} \\ \hline
Icecube Sensitivity & --- & $4 \times 10^{-9}$ & $4 \times 10^{-9}$ & \cite{ri05} \\
\hline
\end{tabular}
\end{center}
\end{table*}

Table 2 shows a comparison of the upper limit on the flux from
starburst galaxies given here with the atmospheric neutrino flux and
with approximate model predictions of neutrino fluxes \gray\ bursts
(GRB) and active galactic nuclei (AGN) along with detector array
sensitivities. It can be seen from this table that at 100 TeV none of the
extragalactic sources proposed will dominate over the atmospheric
foreground. The table shows that
the present preliminary upper limit on the diffuse neutrino energy 
flux below 1 PeV from AMANDA-II is 
$\sim8.8 \times 10^{-8}$ GeV cm$^{-2}$ s$^{-1}$ sr$^{-1}$ 
in the 10 TeV 
to 1 PeV energy range~\cite{hi06}. The full {\it Icecube} detector
array is expected to push down to a sensitivity of 
$\sim 10^{-9}$ GeV cm$^{-2}$ s$^{-1}$ sr$^{-1}$ in the
energy range 100 TeV $< E_{\nu} <$ 100 PeV after several years of
observation.\footnote{F. Halzen, private communication}
Under the extreme assumption that the primary cosmic ray spectra 
in all starburst galaxies are as hard as $E^{-2}$ up to energies 
$\cal{O}$(10 PeV), PeV neutrinos from starburst galaxies may be 
detectable just above the projected sensitivity of {\it Icecube}.\footnote{Even
if we make a second extreme assumption that 100\% of the IR galaxies
at redshifts greater than 1.2 are starburst galaxies, we would still
predict a neutrino flux $< 3 \times 10^{-8}$ GeV cm$^{-2}$ s$^{-1}$ sr$^{-1}$.}
However, as can be seen from Table 2, above 1 PeV the 
AGN models predict fluxes which will be significantly larger than  
the the atmospheric foreground (expected to be $< 3 \times 10^{-9})$ 
GeV cm$^{-2}$ s$^{-1}$ sr$^{-1}$ at 1 PeV), as well as the fluxes from 
\gray\ bursts (GRB) and starburst galaxies.

\section{Observability of Diffuse \grays\ from Starburst Galaxies}

In a follow-up paper to LW, Thompson {\it \etal} ~\cite{th06}
have estimated the contribution of $\pi^{0}$-decay \grays\ from
starburst galaxies to the 
observed \gray\ background in the GeV energy range. Using an $E^{-2}$ 
primary spectrum they get estimates of $E^2\Phi(E)$ fluxes for both 
\grays\ and neutrinos of $3 \times 10^{-7}$ GeV cm$^{-2}$ s$^{-1}$ sr$^{-1}$.
This is a factor of 3 larger than the neutrino flux of LW because they
estimate that all of the electrons in these galaxies are from
$\pi^{\pm}$ decay and they
make the further hypothesis that a significant fraction 
the synchrotron radiating electrons which emit at 1.4 GHz
lose energy by processes other
than synchotron radiation ({\it viz.}, bremsstrahlung and
ionization in dense gas clouds.) Should the diffuse differential
neutrino particle spectrum continue $\propto E^{-2}$ up to 10 TeV and
above, the flux predicted in Ref. \cite{th06} would be more than a factor 
of 3 above the AMANDA-II limit, although no such claim is made in
Ref. ~\cite{th06}.

Our estimated \gray\ flux for the
same $E^{-2}$ primary spectrum is $\sim 2 \times 10^{-8}$ GeV 
cm$^{-2}$ s$^{-1}$ sr$^{-1}$ for \gray\ energies less than $\sim 10$ GeV.
This is $\cal{O}$$(10^{-2})$ of the observed flux of $\sim 1.4 \times
10^{-6}$ GeV cm$^{-2}$ s$^{-1}$ sr$^{-1}$ determined by the {\it EGRET}
group~\cite{sr98}. Above $\sim 10$ GeV the background spectrum will
steepen owing to absorption from pair production interactions with
the extragalactic ultraviolet background radiation~\cite{ss96}.
It should be noted in this context that Stecker and Salamon have shown 
that the bulk of the observed background can be
produced by unresolved blazars~\cite{ss96}.  Thus, the contribution to the 
diffuse \gray\ background from starburst galaxies should be unobservable. 
In this context, we note that
almost all of the observed extragalactic GeV \gray\ sources are blazars; 
no starburst galaxies have been observed.
\section*{Acknowlegdments}

The author would like to thank Eliot Quataert for his comments and
for bringing Ref.~\cite{th06} to his attention.
This work was supported by NASA Grant ATP03-0000-0057.


\begin{thebibliography}{999}

\bibitem{st77} Stecker, F. W. 1977, {\it \apj} {\bf 212}, 60.

\bibitem{st79} Stecker, F. W. 1979, {\it \apj} {\bf 228}, 919.

\bibitem{st75} Stecker, F. W. 1975, {\it \prl} {\bf 35}, 188.

\bibitem{st76} Stecker, F. W. 1976, {\it Nature} {\bf 260}, 412.

\bibitem{to04} Torres, D. F. 2004, {\it \apj} {\bf 617}, 966.

\bibitem{to05} Domingo-Santamar\'{i}a, E. and Torres, D. F. 2005,
{\it Astron. and Astrophys.} {\bf 444}, 403. 

\bibitem{bl99} Blom, J. J.{\it \etal}~ 1999, {\it \apj} {\bf 516}, 744.

\bibitem{lw06} Loeb, A. and Waxman, E. {\it JCAP} 05(2006)003.

\bibitem{ha06} Halzen, F. 2006, \epr 0602132.

\bibitem{th06} Thompson, T. A., Quataert, E. and Waxman, E. 2006,
\epr 0606665.

\bibitem{yu01} Yun, M. S., Reddy, N. A. and Condon, J. J. 2001, {\it \apj}
{\bf 554}, 803.

\bibitem{la05} Lagache,G., Dole, H. and Puget, J.-L. 2005, 
in {\it The Fabulous Destiny of Galaxies: Bridging Past and Present}, in press,
\epr 0509556. 
\bibitem{er06} Erb, D. K., {\it \etal} 2006, {\it \apj} {\bf 646}, 107.

\bibitem{ho03} Hoopes, C. G. {\it \etal}~ 2003 , {\it \apj} {\bf 596}, L175.

\bibitem{bo06} Bouchta, A. 2006, \epr 0606235.

\bibitem{ga05} Gaisser, T. K. 2005, {\it Phys. Scripta} {\bf T121}, 51.

\bibitem{hi06} Hill, G.C. (for the IceCube Collaboration) 2006, 
paper presented at the {\it Neutrino 2006 Intl. Conf.}, Santa Fe, NM.

\bibitem{st05} Stecker, F. W. 2005, {\it Phys. Rev. D} {\bf 72} 107301.

\bibitem{mpr} Mannheim, K., Protheroe, R. J. and Rachen, J. P. 2001,
{\it Phys. Rev. D} {\bf 63} 023003.

\bibitem{wb} Waxman, E. and Bahcall, J. N. 1997, {\it \prl} {\bf 78}, 2292.

\bibitem{ri05}  The Icecube Collaboration (M. Ribordy {\it \etal}), 
\epr 0509322.

\bibitem{sr98} Sreekumar, P. {\it et al} 1998, {\it \apj} {\bf 494}, 523. 

\bibitem{ss96} Stecker, F. W. and Salamon, M. H. 1996, {\it \apj} {\bf 464},
600.

\end{thebibliography}
\end{document}